# Evolution of Interface Magnetism in Fe/Alq$_3$ Bilayer Structure; Thickness-Dependent Interface Resolved Studies Under X-Ray Standing Wave

Avinash Ganesh Khanderao[1,2], V R Reddy[1], Ilya Sergueev[3], and Dileep Kumar[1 a)]

[1]UGC-DAE Consortium for Scientific Research, Khandwa Road, Indore-452001, India.
[2] Department of Physics, Jagadamba Mahavidyalaya, Achalpur City-444806, India.
[3] Deutsches Elektronen-Synchrotron (DESY), Notkestrasse 85, D-22607 Hamburg, Germany.

a) Corresponding author: dkumar@csr.res.in

## Abstract

In the present work, interfacial magnetism at metal organic interface is probed using an isotope sensitive interface resolved nuclear resonance scattering technique which is made depth selective under x-rays standing wave conditions. Using GIWAXS and GINRS measurements, this study evidences the presence of symmetry-based PMA which appears at a lower thickness of Fe having distortion in cubic symmetry and disappears at a higher thickness of Fe as its cubic symmetry retains. The non-zero value of quadrupole splitting evidences the strain at the interfacial region which on increasing thickness of Fe relaxes. The diffusion of Fe is traced using XRF and NRR, deep penetration of Fe in Alq3 layer due to soft nature of the organic film is obtained. This thickness-dependent study enables us to understand the magnetic behavior of buried ferromagnetic metal in the vicinity of organic molecules.

# Introduction

Organic electronics successfully motivated the scientific community over the last few years. This journey of organic electronics began with the occurrence of conducting polymers in 1976 [1]. Organic semiconductor (OSC), on the other hand, was successfully adopted in magnetoresistive tunneling devices [2–4]. The presence of conjugation in OSC with intramolecular covalent carbon-carbon bonding results in enhanced optical properties, whereas intermolecular interaction controls transport properties [5]. Organic films are highly disordered and can be deposited easily using spin coating and thermal evaporation techniques [6]. The field of organic spintronics is emerged due to low spin-orbit coupling, high spin relaxation time and weak hyperfine interaction. All these properties result in the retention of the spin polarization of the carrier for a long time, which allows the manipulation of the spin [7–10]. Since high Spin-orbit and hyperfine interactions are two of the main causes for spins to lose their orientation, this put forward OSC to be used in organic spin valve (OSV) structures [11,12]. OSV are devices that use an organic spacer layer between two ferromagnetic electrodes. It is the spacer thickness that determines the carrier transport [4,13–15]. The failure of OSV is the diffusion of the top electrode into the soft organic layer resulting in the thinning of the effective spacer layer [16,17]. The diffusion in some cases is so high that it forms a conductive pathway to both electrodes [18,19]. Or the effective spacer layer was so thin that the electrons preferred to tunnel through this thinner region rather than hopping through the space [20]. In order to improve the effectivity of OSV, Y. Zhan et al. inserted a barrier layer of $Al_2O_3$ which prevents the high diffusion of upper electrodes [21–23].

In view of this fact, the study of the growth and magnetic behaviour of top electrodes in the vicinity of OSCs is the priority of the researchers since how effective the OSV will be is not realized experimentally to date. Interfaces are crucial in defining and tuning organic devices such as electronic devices based on organic materials such as light-emitting diodes, thin-film transistors, or solar cells. In all such devices, interfaces play a key role in modelling the device properties and their functions [24–28]. The magnetic properties and behavior of metal at the interface is vital in successful working of OSV. Spin transport, Magnetic Spin alignment, hyperfine field at the interface where charge transfer takes place is very imperative to study. Nevertheless, the limitation to most of the available magnetic measurement techniques makes it difficult to study interfacial magnetism unambiguously and thoroughly. The conventional depth profiling techniques, having depth resolution of several nanometers like RBS, SIMS, XPS are unsuitable for organic soft films [29–31]. The high energy ions used in these techniques could result in intermixing and significantly modify the interface. Most of the magnetic measurement techniques provide magnetic information as a whole of the sample, and hence either do not have sufficient depth resolution so as to resolve the interfaces or may not be probing the actual interfaces [25]. X-rays can be made available to probe the interface effectively by confining the X-rays intensity in a specific region along the depth using X-rays standing wave (XSW) generated for waveguide structure [32–34]. In some previous study [25,35–38], the effectiveness of XSW for depth profiling of the interface using the grazing incidence nuclear forward scattering technique is demonstrated. By varying incidence angles, magnetic information was gained from the intended region. It is well studied that magnetic moments are sensitive to their local environments and magnetic field varies from bulk to interfacial region. Also, in case of organic spin valve structure due to diffusion of upper metal into soft organic material, knowledge of how metal behaves magnetically at the buried interface will greatly advances the future OSV and related devices[23,39,40]. As different thicknesses of metal films show different magnetic behaviour, since, these properties were depending on local chemical environment. Magnetic behaviour at organic interface that evolved layer by layer could lead to become a missing bridge for further development of organic electronical devices. Therefore, a thickness dependent magnetic investigation can reveal how metal atom act near organic molecules with different thicknesses. Present work deals with thickness dependent evolution of magnetism at metal organic interface using interface resolved isotope sensitive grazing incidence nuclear resonance scattering (GI-NRS) under XSW conditions which will help in understanding the magnetic behavior of the Fe layer of different thickness at the metal-organic interface and at the buried part, which is essential for the OSV devices functioning.

## Experimental section

The thin film of organic semiconductor Tris(8-hydroxyquinoline) aluminium ($Alq_3$) was deposited using thermal evaporation (TE) in the high vacuum chamber at the base vacuum of $1\times10^{-6}$ mbar [41]. The $Alq_3$ layers were sublimated from a commercial source (sublimated grade, >99% pure, supplied by Sigma Aldrich) at a constant rate of 0.02 nm s$^{-1}$ using the TE technique. During the deposition of $Alq_3$, the crucible temperature was monitored and did not vary by more than ±1°C. Magnetic properties of the films were detected using magneto-optic Kerr effect (MOKE) hysteresis loops both in the longitudinal configurations. NRS is isotope sensitive technique, therefore, six sets of samples Si/Pt (400Å)/$Alq_3$(500 Å)/$^{57}$Fe(X) and (X=20,40,60,80,100 and 120Å) respectively, were also prepared by depositing the Fe isotope - $Fe^{57}$ (95% enriched) marker layer at the interface using electron beam evaporation technique under UHV condition. This is done to increase interface selectivity for the NRS technique. As $^{57}$Fe atoms are expected to diffuse deep into the $Alq_3$ later, resonant $^{57}$Fe nuclei per unit volume drastically decrease. A few angstroms thick $^{57}$Fe layer may spread over up to several nanometres in the $Alq_3$ layer (very less density of $^{57}$Fe resonant nuclei). Therefore, along with depth selectivity in GI-NRS measurement, the enhanced contribution from diffused part is achieved through coinciding the antinode part of XSW with diffused layer by making a waveguide structure to generate XSW. For this purpose, $Alq_3/^{57}Fe$ structure was deposited on a high dense Pt buffer layer. The GI-NRS measurements were carried out at the nuclear resonance beamline at P01 beamline, PETRA III, DESY using photon beam energy of 14.41 keV to excite the $^{57}Fe$ nuclei in the sample [42]. During measurements, the synchrotron was operating in the 40-bunch mode with a bunch separation of 192 ns. The detector used in the experiment was an avalanche photodiode, which has a time resolution of ~1 ns.

## Results and discussions

A set of as-prepared samples was characterized using XRR measurements as a function of scattering vector $q_x$ (increasing incident angle) to get information on the sample structures such as thickness, roughness, and electron densities of the individual layer in the bilayer structure. Figure 1 (a-f) shows the XRR measurements for all six samples, where the solid line shows the best fit to the data. Here reflected X-ray intensities from all the samples are plotted with the magnitude of the momentum transfer vector $q= 4\pi \sin\theta/ \lambda$, where $\theta$ and $\lambda$ are x-rays incident angle and wavelength, respectively. From fig. 1(a-f), initial dips in XRR patterns before the critical angle of Pt ($0.08Å^{-2}$) suggest the presence of X-rays standing wave in all structures. The x-ray reflectivity data are fitted using Parratt's formalism. The final structure obtained from XRR analysis is Si/Pt (400Å)/$Alq_3$(500Å)/$^{57}Fe$(X) and (X=20,40,60,80,100 and 120Å) respectively to get qualitative information about thickness, roughness and density depth profile, XRR curves are fitted using Parratt's formalism [44]. Extracted structural depth profiles of both samples are given in fig. 2. It may be noted that to get the best fit for the XRR experimental data, it is found necessary to divide the Fe layer into diffused layers at the interface with reduced electronic densities. It suggests that the Fe layer diffused deep into $Alq_3$ [44,45]. The SLD profile obtained from Parratt's formulism reveals electron density variation along the depth. This SLD profile provides us an insight view of the thin film structure and it has been observed that the Fe layer penetrated deep inside the $Alq_3$ layer as Fe thickness increases, as shown in figure 2(a-f).

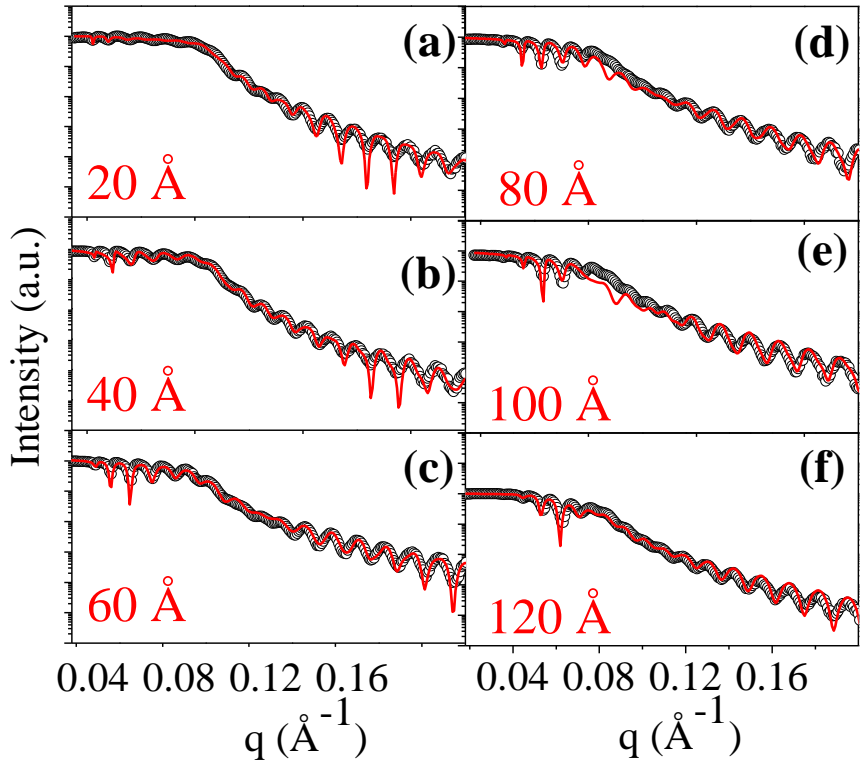

***Figure 1:*** *(a-f) shows the X-ray reflectivity curves for $Si/Pt^{400}/Alq_3^{500}/Fe^{(x=20,40.60,80,100,120Å)}$ respectively. The continuous line shows the best fit to the graph. The clear deeps can be seen before the critical angle of Pt.*

Further, the magnetic behavior of the prepared samples has been studied using MOKE magnetometry measurements in longitudinal geometry. The MOKE hysteresis loops for $Si/Pt^{400Å}/Alq_3^{500Å}/Fe^{(x=20,40.60,80,100,120Å)}$ is as shown in figure 3 (a-f) respectively.

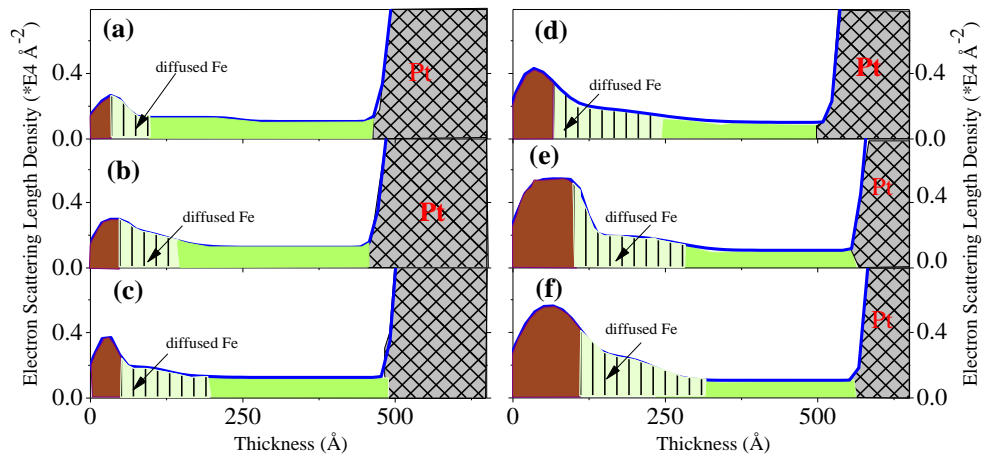

***Figure 2:*** *(a-f) represents the electron density profile as a function of depth of the samples $Si/Pt^{400}/Alq_3^{500}/Fe^{(x=20,40.60,80,100,120Å)}$ respectively.*

It has been found that, up to the thickness of 60 Å MOKE hysteresis loop does not appear but starts appearing from the Fe thickness of 80Å onwards.

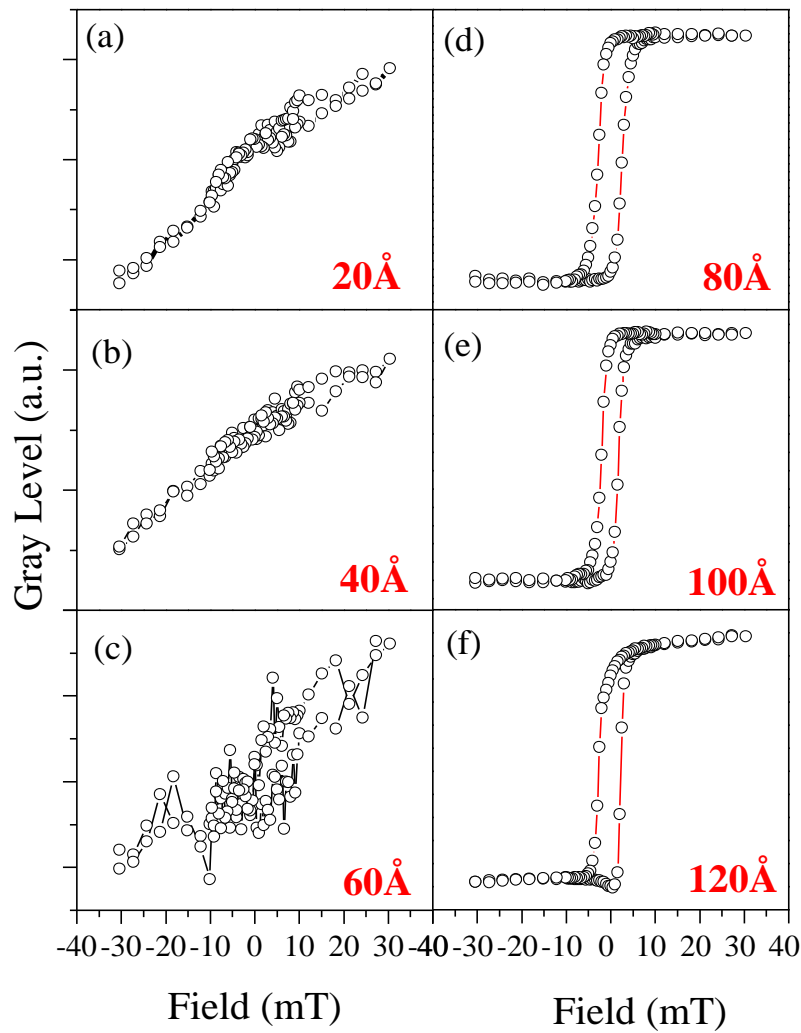

*Figure 3: (a-f) MOKE loop as a function of thickness of Fe on* Alq$_3$ *using Kerr microscopy.*

Further, MOKE magnetometry revels the variation in the coercivity where, the coercivity value is reduces for the Fe thickness of 100 Å than that of 80 Å and on further increase in the Fe thickness, coercivity increases, this behaviour is attributed to the release of the stress of the film generated at lower thickness. And as the thickness increases, all the bigger clusters joined together and after that the stable magnetic signal is obtained for higher magnetic thickness [46]. The magnetic information obtained from the MOKE measurements mainly concerns the signal from the whole sample and has limitation for the laser penetration at higher thickness. So, MOKE measurements could not be employed for the interfacial or buried magnetic measurements effectively. As Fe atoms penetrates deep in the soft organic layer Alq$_3$, the magnetic behaviour of Fe in the vicinity of organic molecules is crucial to understand for better working of organic spin valve structures.

Now, to investigate morphology of various thickness, the growth pattern of the samples has been studied using GIWAXS, a photon energy of 14.4 keV with beam size of 48 × 5 (µm)2 (horizontal to vertical ratio) was used to illuminate the sample at a grazing-incident angle, larger than the critical angle for total external reflection, at which X-rays penetrate the sample. The experimental geometry is as shown in fig 4(a). For this study, two samples, one form non-magnetic region (20 Å) and other from magnetic region (80 Å) has been selected. Static GIWAXS measurements performed on two samples is shown in fig 4 (b) and (c), the peaks corresponding to Fe are marked in the figure. The one-dimensional scattering curves obtained by radial integration from the GIWAXS pattern using DPDAK[47] software is shown in fig 4 (d) and (e) respectively for 4 nm and 8 nm samples. GIWAXS data reveals that for 20A sample, continuous ring associated with the Fe(110) plane formed with random orientation, with a more prominent preference for out-of-plane orientation. Whereas, for 80A sample shows discontinuous rings with strong texturing along\ in-plane orientation.

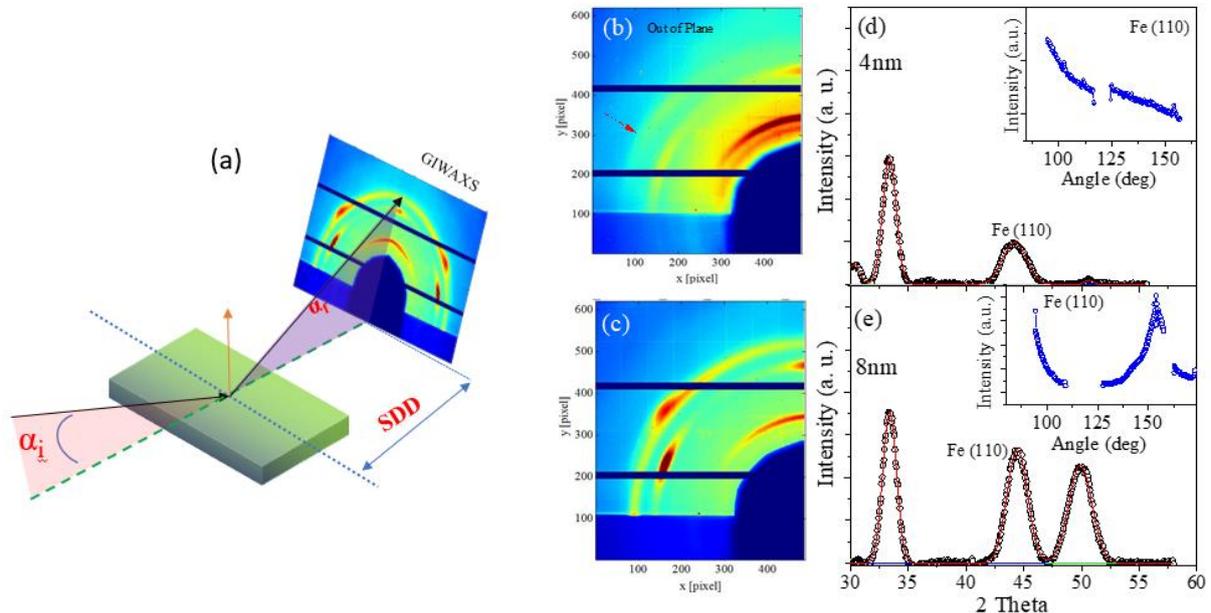

**Figure 4:** (a) Experimental geometry of GIWAXS measurement. 2D GIWAXS image for 4nm and 8nm film is shown in (b) and (c) respectively. The 1D extraction of 4 nm and 8nm film is shown in (d) and (e). Inset to (d) and (e) shows the azimuthal extraction of Fe (110).

From MOKE ang GIWAXS measurements it has been clear that all samples have different magnetic and growth properties, hence, to corelate this magnetic and morphological properties and to study the evolution of magnetism (Fe$_{diffused}$) we have employed nuclear resonant scattering of synchrotron radiation at the 14.4 keV resonance of $^{57}$Fe. GINRS techniques is time analogue to Mossbauer spectroscopy. The simultaneous excitation of the hyperfine-split nuclear energy levels caused by a radiation pulse generates quantum beats in the temporal evolution of the subsequent nuclear decay signal. An accurate assessment of the magnitude and direction of magnetic fields in the sample is possible through examination of this beat pattern. Along with the isotope sensitivity GINRS technique can be made depth selective under the x-ray standing wave condition [25,35,36]. For this purpose, we have used $^{57}$Fe in the as prepared samples where different set of samples are having planer waveguide structure such that each sample should satisfy the x-ray standing wave condition. In view of fact, we have used Pt as buffer layer which will act as a mirror to x-rays to satisfy total internal reflection. Hence the final structures that we have deposited are Si/Pt/Alq$_3$/$^{57}$Fe$^{(x=20,40,60,80,100 \text{ and } 120\text{Å})}$. The final structure comprises a low dense organic semiconductor layer of Alq$_3$ which is sandwiched between high dense Pt layer and Fe layer. Here, Alq$_3$ layer act as a guiding layer. To achieve the depth selectivity, XRF measurements were performed simultaneously with GINRS measurements. The setup for the measurements is shown in fig 4. Where Fluorescence detector is placed normally to the film plane. The obtained XRF data is fitted with the Parratt's formulism simultaneously with XRR to ensure the best fit model. Using this Parratt's model, a 2D contour plot is simulated.

This 2D simulation is the distribution of x-ray filed intensity inside the waveguide structure. Where, the node-antinode pattern obtained is used to select the depth region for GI-NRS analysis. Figure 5 (a-f), shows the consolidated XRR, XRF and 2d counter plot respectively for Si/Pt/Alq$_3$/Fe$^{(x=20, 40, 60, 80, 100 \text{ and } 120\text{Å})}$. Excitation of XSW modes is represented using the dotted lines which corresponding to the different antinodes. The NRR measurements again confirms the presence of XSW as shown in figure 6 (a). Using 2D simulation from fig 5, a region of interest is selected for the NFS measurement by selecting an incident angle. Here at particular angle, NFS data will generate a beating pattern which is mainly associated with the selected region (buried layer). The NFS time spectra for different samples were recorded at an incident angle of 0.150°, 0.197°, 0.140°, 0.204°, 0.196° and 0.165° for Si/Pt/Alq$_3$/Fe$^x$ (x=20,40,60,80,100 and 100 Å) respectively is as shown in fig. 6 (b).

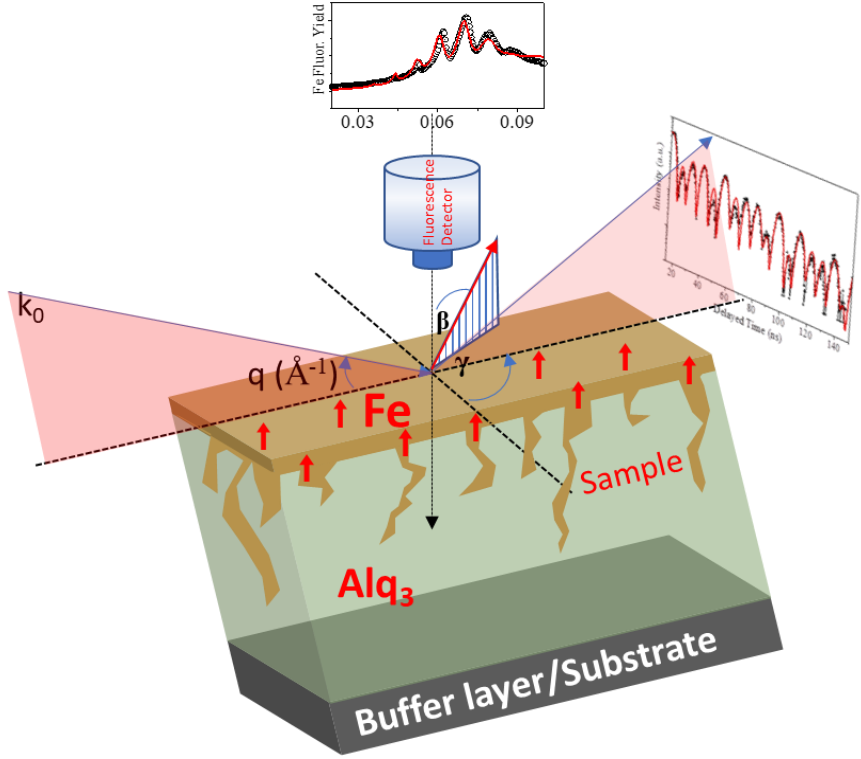

*Figure 4*: Experimental set up used for NFS and XRF simultaneously

The NFS data is then fitted using REFTIM software [48] . To get a best fit to the data, an interfacial layer having reduced hyperfine field was introduced. In order to obtain more precise information about the hyperfine parameter, the simultaneous fitting of both NRR and NFS time spectra was performed.

From Fig. 6 (c), one may note that up to a thickness of 60 Å, the hyperfine field value is near to zero, hence suggesting the reduced magnetism at the initial thickness. The reason for the reduced magnetism can be attributed to the superparamagnetic relaxation of islands [36]. This superparamagnetic relaxation is seen when the magnetic anisotropy energy of a single-domain particle and thermal energy are comparable. Since the thermal fluctuations make the magnetization of the particle fluctuate along different easy magnetization directions. Further at a thickness of 80 Å, the film exhibits a finite hyperfine field (HF) comparable to bulk Fe. Further increase in thickness, HF increases linearly and slowly. It may be noted that once macroscopic islands form as a result of percolation transition, the relaxation of their magnetic moments will slow down, resulting in a finite hyperfine field. Thus, GINRS measurements suggest that percolation transition takes place around a thickness of 60 Å [36]. At a thickness of 100 Å, the film exhibits a hyperfine field value of 33 T, which corresponds to bulk Fe.

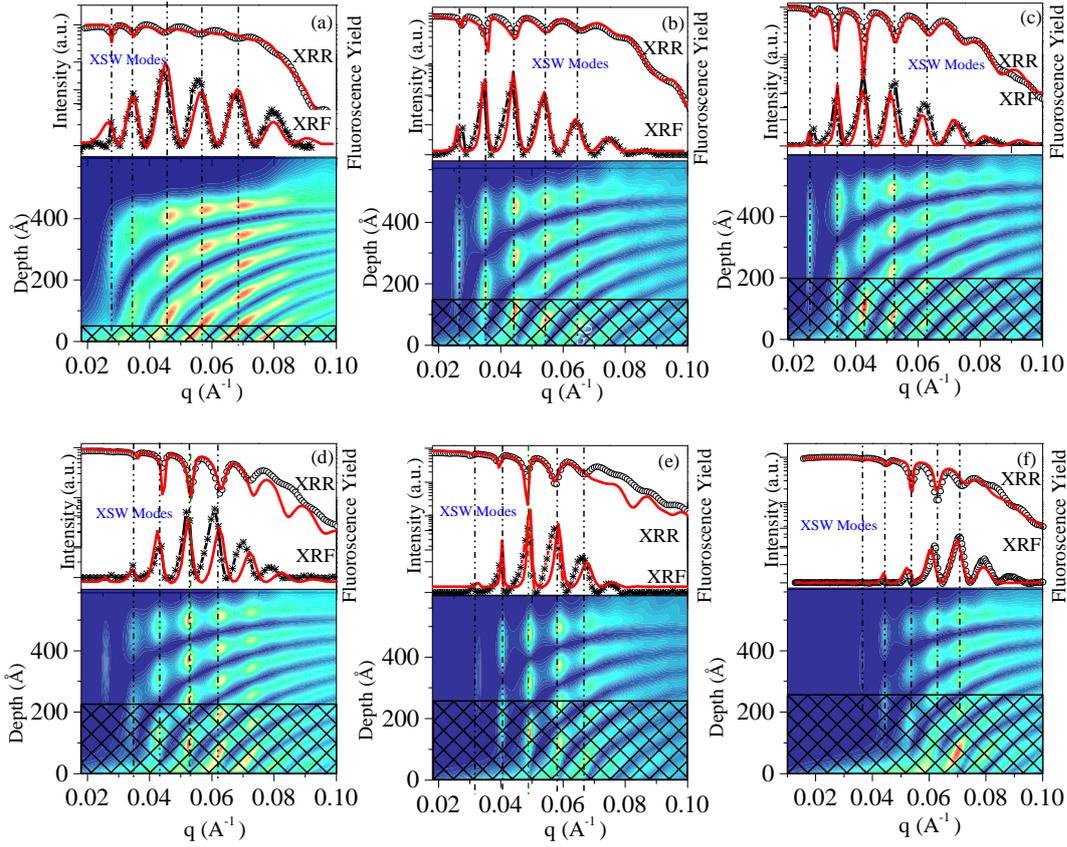

**Figure 5**: *(a-f) shows the consolidated XRR, XRF and 2d counter plot respectively for Si/Pt/Alq$_3$/Fe$^{(x= 20, 40, 60, 80, 100 \text{ and } 120 Å)}$. Excitation of XSW modes is represented using the dotted lines which corresponding to the different TE modes.*

The finite value of QS suggests a distortion from cubic structure, as for a cubic symmetry the QS is expected to be zero. Also, this thickness dependency of QS is related to strain generated in the film during growth. There are several ways to generate strain. For metal-organic interfacial region, the strain caused is due to the large clusters of Fe formed before the percolation. Also, this variation of QS depends on the Fe environment which is mainly an organic molecule in the initial state of deposition. The angle β between the film surface normal and the direction of hyperfine field is about 0° for film thickness up to 60 Å Which suggests the presence of small perpendicular magnetic anisotropy (PMA) during the initial thickness. However, for a film thickness of 80 Å and above, β goes to 90°, this signifies that the magnetic moment becomes in-plane to the film surface. Hence, at this thickness, the PMA disappears as shown in the schematic in fig 7. It is interesting to note that the film exhibits a PMA as long as its structure deviates from cubic symmetry, as evidenced by a finite value of QSs.

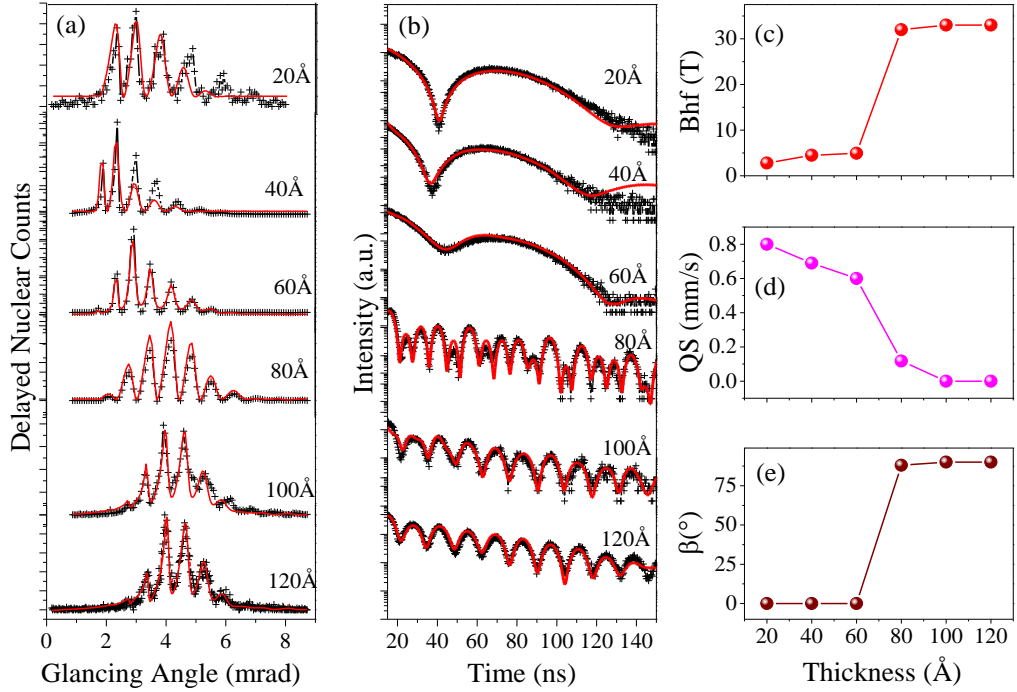

***Figure 6***: *(a) NRR spectra for Si/Pt/*Alq$_3$*/Fe$^x$ (x= 20,40,60,80,100 and 120Å) respectively. (b)NFS time spectra for Si/Pt/*Alq$_3$*/Fe$^x$ (x= 20,40,60,80,100 and 120Å) respectively. (c)Variation of hyperfine field (Bhf), (d) quadrupole splitting (QS), and (e) average angle of atomic spins with respect to surface normal (β°) as a function of film thickness*

As the QS goes to zero and the film attains cubic symmetry, the PMA also disappears. This observation provides clear evidence that structural distortion in thin film is at least partly responsible for the observed PMA. The pictorial representation for the spin orientation is shown in fig 7. Thus, present GINRS measurements suggest that the observed PMA in Fe on Alq$_3$ is mainly due to an interfacial phenomenon that disappears for higher thickness.

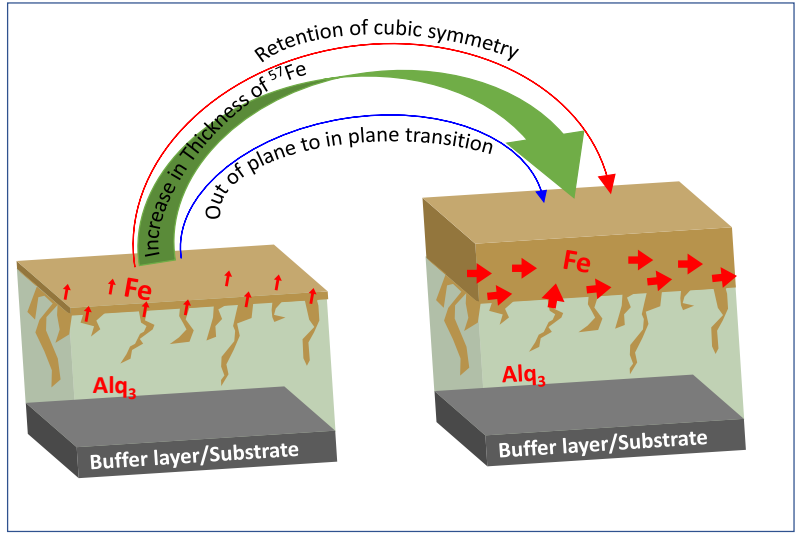

*Figure 7*: Schematic showing the presence of PMA at lower thickness while Fe moments aligned in-plane with increased thickness.

# Conclusion

The present work provides insight into the evolution of interface magnetism at the metal-organic interface (upper interface) and effectively probed buried magnetism for the metal-organic framework. Decisive information at the metal-organic interface is observed at lower thickness using an isotope sensitive interface resolved nuclear resonance scattering technique which is made depth selective under x-rays standing wave condition to effectively probe the interfacial magnetism. The diffusion of Fe is traced using X-ray fluorescence and nuclear fluorescence. Here using GINRS and GIWAXS, we witness the presence of symmetry-based PMA which disappears at a higher thickness of Fe as cubic symmetry is retained. The non-zero value of quadrupole splitting evidences the strain at the interfacial region which on increasing thickness of Fe relaxes. The presence of a dead layer and super-paramagnetism is confirmed using both MOKE microscopy and GINRS.

# Acknowledgements

Portions of this research were carried out at the light source, PETRA III of DESY, a member of Helmholtz Association (HGF). Financial support from the Department of Science and Technology (Government of India) (Proposal No. I-20180885) provided within the framework of the India@DESY collaboration is gratefully acknowledged.